\documentclass[aps,prl,twocolumn,showpacs,preprintnumbers,amsmath,amssymb,floatfix]{revtex4}

\usepackage{graphicx}
\usepackage{amsmath}

\newcommand{\an}{\hat{a}}
\newcommand{\bn}{\hat{b}}
\newcommand{\ad}{\hat{a}^\dagger}
\newcommand{\bd}{\hat{b}^\dagger}

\begin{document}

\title{Generation of Spatially Broadband Twin Beams For Quantum Imaging}

\author{V. Boyer}
\affiliation{Joint Quantum Institute, National Institute of
Standards and Technology, and University of Maryland, Gaithersburg, 
Maryland 20899, USA}
\author{A. M. Marino}
\affiliation{Joint Quantum Institute, National Institute of
Standards and Technology, and University of Maryland, Gaithersburg, 
Maryland 20899, USA}
\author{P. D. Lett}
\affiliation{Joint Quantum Institute, National Institute of
Standards and Technology, and University of Maryland, Gaithersburg, 
Maryland 20899, USA}

\date{\today}

\begin{abstract}

    We generate spatially multimode twin beams using 4-wave mixing in
    a hot atomic vapor in a phase-insensitive traveling-wave amplifier
    configuration.  The far-field coherence area measured at 3.5 MHz
    is shown to be much smaller than the angular bandwidth of the
    process and bright twin images with independently
    quantum-correlated sub-areas can be 
    generated with little distortion.  The available transverse
    degrees of freedom form a high-dimensional Hilbert space which we use
    to produce quantum-correlated twin beams with finite orbital
    angular momentum.  

\end{abstract}


\maketitle

Quantum imaging is concerned with extending the realm of nonclassical
optical effects from the time and frequency domain to the spatial degrees
of freedom~\cite{kolobov_spatial_1999}.  Spatially multimode squeezed
light holds the promise of improved optical
resolution~\cite{kolobov_quantum_2000}, quantum holographic
teleportation~\cite{sokolov_quantum_2001}, and parallel quantum
information encoding~\cite{lassen_tools_2007}.  Recently, it enabled
the detection of transverse beam displacements smaller than the
standard quantum limit (SQL)~\cite{treps_quantum_2003}, and noiseless
image amplification has been
demonstrated~\cite{choi_noiseless_1999,mosset_spatially_2005}.
Previous approaches to the generation of nonclassical multimode light
have focused on parametric down-conversion in a $\chi^{(2)}$ crystal,
following two basic configurations.  In the first one, a
traveling-wave parametric down-converter is shown to be intrinsically
multimode~\cite{jedrkiewicz_detection_2004}.  Pulsed operation is
required to compensate for the weakness of the nonlinear coefficient.
In the second approach, an optical parametric oscillator operates
continuously inside a cavity with degenerate transverse
modes~\cite{martinelli_experimental_2003}.  This device is difficult
to control above threshold because different transverse modes tend to be
strongly coupled, and an alternative is to tune the cavity to be
resonant with a single higher order transverse mode.  Orthogonal modes
separately prepared in a nonclassical state can then be combined with
a specially crafted beamsplitter~\cite{lassen_tools_2007}.

Following the recent development of a 4-wave mixing (4WM) amplifier
operating deep in the quantum
regime~\cite{mccormick_strong_2007-1,mccormick_strong_2007}, we
present a different approach based on 4WM in an atomic
vapor~\cite{kumar_degenerate_1994}.  It combines the advantages of
both of the above approaches by exploiting a large single-pass gain.
In the absence of a cavity, no mode selection occurs and spatially
multimode twin beams can be generated in the high-intensity regime,
with little optical aberrations.  As a result, unprecedented levels of
spatial quantum correlations between the twin beam intensities are
recorded.

\begin{figure}[htb]
    \begin{center}
	\includegraphics[width=.9\linewidth]{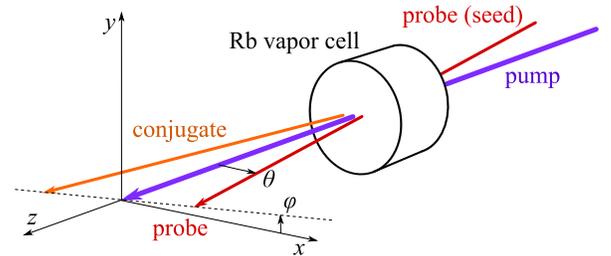}
    \end{center}
    \caption{(color online). Setup geometry.}
    \label{fig:exp}
\end{figure}

The setup has been described in
Ref.~\cite{mccormick_strong_2007-1} and is sketched in
Fig.~\ref{fig:exp}. A weak probe beam intersects a strong (420~mW)
pump beam inside a 12 mm-long $^{85}$Rb cell at a small polar angle
$\theta$ and with an azimuthal angle $\varphi$.  The pump has a
550~$\mu$m waist ($1/e^2$ radius) and the beams overlap over the full
length of the cell.  The pump and the probe, with angular frequencies
$\omega_0$ and $\omega_p < \omega_0$ respectively, 
are tuned $\approx 800$ MHz to
the blue of the $D1$ line at 795 nm and are resonant with a 2-photon
Raman transition between the two electronic ground states $F = 2$ and
$F = 3$, which are separated by 3~GHz.  The coupling between the light
fields and the atomic levels follows a double-lambda configuration and
gives birth to a 4WM process which converts two photons from the pump
into one probe photon and one conjugate photon at the frequency
$\omega_c = 2\omega_0 - \omega_p$~\cite{mccormick_strong_2007-1}.  As
a result, the probe beam is amplified and a conjugate beam emerges at
an angle $\theta$ from the pump, on the other side from the probe
(azimuth $\varphi + \pi$).  The noise on the intensity difference
between the probe and the conjugate is recorded with an amplified
balanced photodetector and analyzed with a spectrum analyzer. The
total detection efficiency is 0.9 and the noise measurements are
corrected only for the detector electronic noise.  The estimated error
on the normalized noise measurements (i.e. the noise level with
respect to the SQL) is less than $\pm 0.2$ dB. The conjugate optical
path is lengthened by almost 10 ns to compensate for propagation
effects~\cite{boyer_ultraslow_2007}.  Distances measured
perpendicularly to the propagation direction ($z$ axis) in the far
field are reported here as divergence angles with respect to the
center of the cell.

The 4WM is a phase-insensitive amplification process which can be
ideally described by the two-photon squeeze operator $\hat S_{ab} =
\exp (s \an \bn - s \ad \bd)$ where $\an$ and $\bn$  are the photon
annihilation operators for the optical modes of the probe and the
conjugate respectively. The (chosen to be real) squeeze parameter $s$ is
related to the gain $G$ by $G = \cosh^2 s$.
The input probe field is a coherent state, and the input
conjugate field is the vacuum.  For large values of $G$ (4--9
in our experiment), this process leads to substantial quantum
intensity correlations between the probe and the conjugate.  For a
Gaussian probe beam of waist half the size of the pump waist in the
medium, the measured probe-conjugate intensity-difference quantum
noise reduction is more than $-8$ dB at
1~MHz~\cite{mccormick_strong_2007}, and squeezing is observed at
frequencies up to 20 MHz.

The first evidence of the multimode behavior of our system stems from
the fact that it can operate for various angles $\theta$ and $\varphi$
of the probe beam.  Given the cylindrical symmetry of the setup, the
absence of dependence on the azimuthal angle $\varphi$ is obvious.  On
the other hand, the angle $\theta$ is set by the phase matching which,
in the case of the forward 4WM geometry used here, is usually a
stringent condition~\cite{kumar_degenerate_1994}.  In our case, two
factors moderate this conclusion.  First, although the phase matching
condition in vacuum would command all the beams to be perfectly
aligned, the appearance of a strong dispersion of the index of
refraction for the probe~\cite{boyer_ultraslow_2007} allows the
existence of a pair of frequencies for the pump and the probe such
that the 4WM gain is maximum for a non-zero $\theta$, around
$\theta_0 = 7$ mrad.  Second, the large gain exhibited by the atomic
medium makes it possible to use a relatively thin medium.  Therefore
there should be a sizeable range of angles $[\theta_0 - \theta_m,\,
\theta_0 + \theta_m]$ for which the dephasing length is longer than
the medium length and for which the beams are quasi-phase matched.  

\begin{figure}[htb]
    \begin{center}
	\includegraphics[height=.9\linewidth,angle=-90]{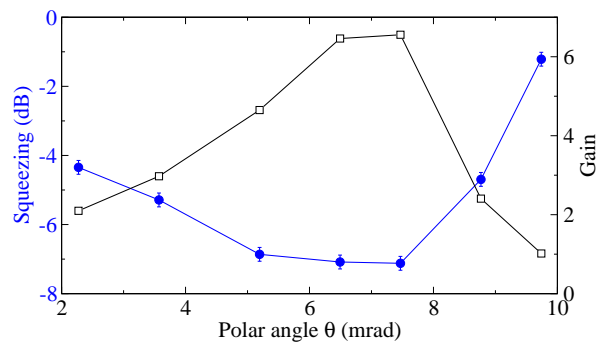}
    \end{center}
    \caption{Gain ({\tiny $\square$}) and quantum noise reduction on
    the intensity difference ({$\bullet$}) as a function of the angle
    $\theta$.}
    \label{fig:acceptance}
\end{figure}

The maximum mismatch angle for which the dephasing length is equal to
the cell length $L$ is $\theta_m \approx \sqrt{\lambda/L} = 8$ mrad.
Figure~\ref{fig:acceptance} shows the variations of the gain and the
intensity-difference squeezing measured at 1~MHz as a function of
$\theta$.  The width of the squeezing dip sets the angular bandwidth
to be $\Delta \theta \approx 8$ mrad.  The sharp decrease in gain and
squeezing at larger angles suggests that the angular bandwidth is
limited in that region by the overlap between the probe and the pump
beams rather than by the phase matching.  

It should be emphasized that the angular bandwidth is determined here
while keeping all the experimental parameters (temperature, pump
intensity, pump and probe detuning) constant.  As a consequence, our
system can be more flexible than systems using cavities having
non-degenerate transverse modes.  Within the angular bandwidth, the
4WM is truly multimode and can generate quantum correlations
simultaneously on a collection of modes without the need of the
multiplexing and demultiplexing described in
Ref.~\cite{lassen_tools_2007}.  

Formally, a multimode state of the light field is a state which cannot
be written as a single occupied optical mode with all the other modes
in a vacuum state~\cite{martinelli_experimental_2003}:
\begin{equation} |\Psi\rangle = |\Psi_0\rangle_0 \otimes |0\rangle_1
\otimes \cdots \otimes |0\rangle_i \otimes \cdots,
\label{eq:singlemode} \end{equation} where $i$ indexes a basis of
optical modes coupled to the medium.  Even when the probe is
seeded with a coherent state, which is a single mode according to
the definition above, and the conjugate is seeded with the vacuum,
the output of the medium is multimode.  Because many transverse
optical modes are coupled to the medium, the 4WM is described by a
collection of squeeze operators $\hat S_{a_i b_i} = \exp (s_i
\an_i \bn_i - s_i \ad_i \bd_i)$, one for each pair of coupled
probe-conjugate modes, described by the photon annihilation
operators $(\an_i,\, \bn_i)$.  These operators transform the probe
and conjugate modes which are initially in a vacuum state into a
two-mode squeezed vacuum state of squeeze parameter $s_i$
containing spontaneously emitted photon pairs.  The output of the
medium when the process is seeded by a coherent state on the probe
is thus a pair of bright twin beams surrounded by orthogonal
two-mode squeezed vacuum modes.  
When the entire transverse spatial extent of the bright twin beams is
collected by the balanced photodetector, the measured intensity noise
is dominated by the two bright modes, and these do not beat with any
of the orthogonal vacuum transverse modes.  In contrast, when the
bright twin beams are partly blocked, some of the transverse vacuum
modes are not orthogonal to the detected area of the bright modes and
beat against them. 
This leads to a
modification of the noise properties of the beams which depends on the
size and the position of the detection area.

The dependence on the detection area suggests a method to discriminate
between single- and multi-modes~\cite{martinelli_experimental_2003}.
The Mandel parameter is defined as $Q = (\langle \Delta N^2\rangle /
\langle N \rangle) - 1$, where $N$ is the photon number operator. It
is the intensity noise normalized to the noise of a coherent state of
same intensity $\langle N \rangle$ (i.e. the SQL), less one.  When a
light field is attenuated with a beamsplitter, $Q$ varies linearly
with the transmitted intensity, and in the limit of null transmission
tends to 0, the value for a coherent state.  As shown in
Ref.~\cite{fabre_quantum_2000}, the $Q$ parameter for a partially
detected single-mode beam varies as if the beam was simply attenuated
so as to give the same detected intensity.  On the contrary, a
multimode field has a $Q$ parameter which is not related to the mean
value of the transmitted field.  For instance, a field composed of
independent beams having the same intensity noise will exhibit the
same $Q$ whatever number of those beams fall onto the photodetector.  

\begin{figure}[htb]
    \begin{center}
	\includegraphics[height=\linewidth,angle=-90]{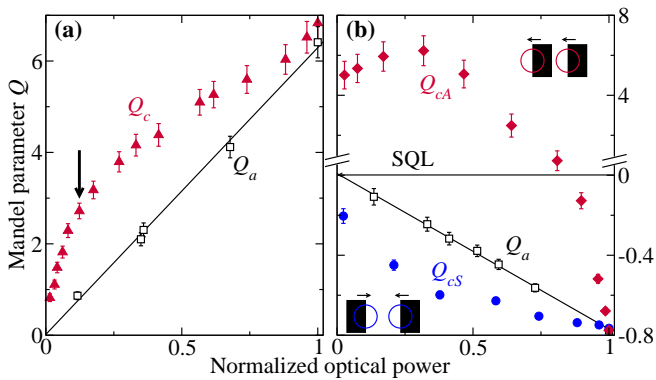}
    \end{center}
    \caption{(color online). Mandel parameter for the probe only (a) and
    for the intensity
    difference (b) as a function of the transmission.  The
    beams are either attenuated ({\tiny $\square$}) or clipped with an
    iris ({\footnotesize $\blacktriangle$}) or with
    edges placed symmetrically ($\bullet$) or anti-symmetrically
    ({\scriptsize $\blacklozenge$}) in the far field.  Note the 
    change of scale on the rightmost $y$ axis at 0.}
    \label{fig:mandelq}
\end{figure}

To confirm the multimode character of an amplified Gaussian probe
(coherent state input), we plot in Fig.~\ref{fig:mandelq}(a) the
$Q$ parameter measured at 1~MHz as a function of the intensity, when
the output probe is either attenuated ($Q_a$) or clipped ($Q_c$) with a
centered iris placed in the far field, at 3 times the pump Rayleigh range
after the cell~%
\footnote{The output
probe alone is not a pure state but a \emph{stochastic} field.  We
will assume that the criteria for the spatially multimode character is
still valid.}.  
For a transmission of one, the output beam displays some excess
noise because of the phase-insensitive amplification process.
As expected, $Q_a$ varies linearly with the transmitted intensity,
while $Q_c$, starting from the full transmission
and reducing the intensity, decreases more slowly than
$Q_a$. This indicates a certain degree of independence of the
various transverse areas with respect to their intensity noise.

We now address the multimode character of the strong quantum
correlations between the twin beams.  To this end, the $Q$ parameter
is now calculated for the difference $N = N_p - N_c$, where $N_p$ and
$N_c$ are the probe and the conjugate photon number operators.
Figure~\ref{fig:mandelq}(b) represents $Q$ when the beams are either
attenuated with a beam splitter ($Q_a$) or clipped along the polar
direction with a straight edge in the far field.  The edges clip the
same fractional amount of light from the probe and the conjugate
symmetrically ($Q_{cS}$) or anti-symmetrically ($Q_{cA}$) with respect
to the pump, as shown in Fig.~\ref{fig:mandelq}(b).  The analyzing
frequency is now 3.5 MHz (see below), and the measured squeezing at a
transmission of 1 is $-6.5$ dB, corresponding to a gain $G = 4.5$.  As
in the single beam case, $Q_a$ varies linearly whereas $Q_{cS}$ and
$Q_{cA}$ have non trivial variations.  The fact that $Q_{cS}$ is
smaller than $Q_a$ shows that the squeezing is spatially multimode and
that the intensity fluctuations are symmetrically correlated with
respect to the pump axis, as one would expect from the transverse
phase matching condition.  Uncorrelated parts of the beam
cross-sections result in excess noise in the intensity difference, as
shown by the behavior of $Q_{cA}$ [note the change of scale in
Fig.~\ref{fig:mandelq}(b)].

The fact that, in both the single beam and the dual beam cases, the
$Q_c$'s do not stay constant and eventually tend to zero in the limit
of a small detection area hints at the existence of an area scale,
called the coherence area~\cite{brambilla_simultaneous_2004}, under
which the light field behaves like a single mode.  This is because the
finite transverse size of the gain area, given by the near-field pump
size, limits the smallest divergence that the coupled probe/conjugate
modes can have to roughly the divergence of the diffraction-limited
pump.  In the single beam experiment, an estimate of the decreasing
intensity at which $Q_c$ starts converging towards $Q_a$, indicated in
Fig.~\ref{fig:mandelq}a by the arrow, corresponds to an aperture size
of the iris of 1.4 mrad.  This value is comparable to the size of the
pump in the far field ($1/e^2$ diameter of 1.0 mrad).

\begin{figure}[htb]
    \begin{center}
	\includegraphics[height=\linewidth,angle=-90]{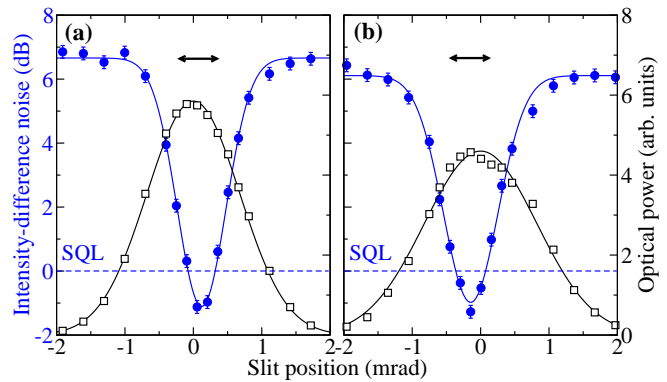}
    \end{center}
    \caption{(color online). Intensity-difference noise ($\bullet$)
    and transmitted probe power ({\tiny $\Box$}) as a function of the
    position of a slit placed on the probe while a similar slit is
    placed at a fixed position on the conjugate.  The slits are
    orientated in the azimuthal (a) or polar (b) direction with
    respect to the pump.  The data is fitted with Gaussians.  The size
    of the slits is indicated by the double arrows.}
    \label{fig:scan}
\end{figure}

Although the variations of $Q_{cS}$ contain the effects of the finite
coherence area, it is difficult to extract a length scale, as factors
like the finite size of the beams or the amount of excess intensity
noise come into play.  To get a more precise estimate of the
length scale which characterizes the spatial intensity correlations
between the probe and conjugate, we select a narrow band of the
conjugate beam cross-section with a slit placed roughly at the center
of the beam and we scan another slit with the same orientation across
the probe beam.  As shown in Fig.~\ref{fig:scan}, the
intensity-difference excess noise (in dB), measured at 3.5 MHz,
displays a dip whose position indicates the location of the probe area
which is correlated with the selected conjugate area.  The size of the
dip gives an estimate of the size $\theta_c$ of the coherence area.
After deconvolution from the effect of both slits, we find
$\theta_c=1.2$ mrad (full width at $1/e^2$) averaged over both
orientations of the slits.  These measurements were made with a gain
of about 4.5, yielding $-6.5$ dB of intensity-difference squeezing for
the full twin beams.  At larger gain, two effects may contribute to an
observed increase of  the size of the coherence area.  First, the
nonlinear dependence of the gain on the pump intensity leads to an
effective  narrowing of the gain
area~\cite{jedrkiewicz_detection_2004}. Second, cross-phase modulation
between the pump and the probe introduces optical aberrations on the
probe.  An obvious way of reducing these effects would be to use a
larger pump beam waist.

A rough estimate of how many independently quantum-correlated pairs of
probe/conjugate optical modes can be generated by the 4WM process can be
made by counting the number of coherence areas that can be fit in
the (solid) angular bandwidth.  That number is $\Delta\theta /
\theta_c \approx 6$ radially and $\pi\theta_0 / \theta_c \approx 18$
in the azimuthal direction, for a total number of modes of about
$100$.  

To demonstrate the multimode capabilities of the system, we
generate twin beams out of non-trivial optical modes.  In a first
experiment, we seed the 4WM process with a probe made out of two spots
(two Gaussian beams) having the same polar angle $\theta$ and
separated by 3 mrad in the azimuthal direction.  The quantum
intensity-difference noise reduction measured when the two spots for
both the amplified probe and the conjugate hit the balanced detector
is $-6.6$ dB, and the conjugated spot pairs measured individually yield
a quantum noise reduction of $-5.9$ dB and $-5.8$ dB (all measured at 3.5
MHz).  At frequencies lower than 3 MHz, we found that, while the
global squeezing is unchanged, the squeezing for each individual
probe/conjugate pair is degraded by the presence of the other
(undetected) pair.  The origin of this coupling between the beam pairs
will be studied elsewhere.  The independence of the conjugated spot
pairs in terms of the correlations at certain frequencies demonstrates
the possibility of using the transverse structure of the twin beams as
a quantum resource for parallel continuous-variable (CV) operations on
the field quadratures.

\begin{figure}[htb]
    \begin{center}
	\vspace{1ex}
	\includegraphics[width=.81\linewidth]{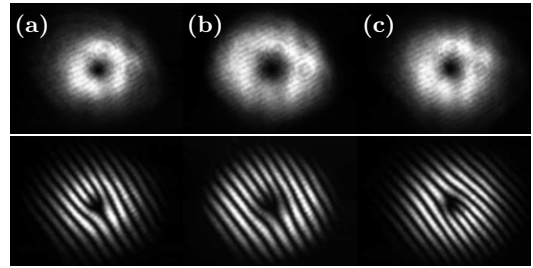}%
    \end{center}
    \caption{Intensity profiles of the beam at the input (upper row)
    and the output (lower row) of the interferometer for: (a) the
    input probe, (b) the output probe, (c) the conjugate.}
    \label{fig:lg}
\end{figure}

In a second experiment, we seed the process with a Laguerre-Gauss (LG)
mode of order 1, and we check the phase properties of each beam by
interfering it with its mirror image, in an optical arrangement
equivalent to the one of Ref.~\cite{harris_laser_1994}.
Figure~\ref{fig:lg} shows the intensity profiles and the
interferograms of the input probe and the output twin beams.  The
fringe count on the interferograms shows that the process produces
twin LG modes of order $\pm 1$.  The measured intensity-difference
squeezing between the output LG modes is $-7.3$~dB, which demonstrates
the possibility of using this system to encode quantum information in
the spatial basis of the orbital angular
momentum~\cite{molina-terriza_twisted_2007}.

In conclusion, we have shown that 4WM in an atomic vapor can generate
highly quantum-correlated twin beams in a large number of transverse
modes simultaneously. This opens the door to CV quantum information
encoding in a high-dimensional Hilbert space.  A natural extension of
this work would be to demonstrate the entanglement of these modes,
both in the phase/intensity~\cite{villar_generation_2005} and in the
displacement/tilt~\cite{delaubert_tem[sub_2006} spaces.  Finally,
direct imaging of the spatial distribution of the fluctuations with a
camera, in the spirit of Ref.~\cite{mosset_spatially_2005}, should
give access to the spatial frequency spectrum of squeezing.


\begin{thebibliography}{19}
\expandafter\ifx\csname natexlab\endcsname\relax\def\natexlab#1{#1}\fi
\expandafter\ifx\csname bibnamefont\endcsname\relax
  \def\bibnamefont#1{#1}\fi
\expandafter\ifx\csname bibfnamefont\endcsname\relax
  \def\bibfnamefont#1{#1}\fi
\expandafter\ifx\csname citenamefont\endcsname\relax
  \def\citenamefont#1{#1}\fi
\expandafter\ifx\csname url\endcsname\relax
  \def\url#1{\texttt{#1}}\fi
\expandafter\ifx\csname urlprefix\endcsname\relax\def\urlprefix{URL }\fi
\providecommand{\bibinfo}[2]{#2}
\providecommand{\eprint}[2][]{\url{#2}}

\bibitem[{\citenamefont{Kolobov}(1999)}]{kolobov_spatial_1999}
\bibinfo{author}{\bibfnamefont{M.~I.} \bibnamefont{Kolobov}},
  \bibinfo{journal}{Rev. Mod. Phys.} \textbf{\bibinfo{volume}{71}},
  \bibinfo{pages}{1539} (\bibinfo{year}{1999}).

\bibitem[{\citenamefont{Kolobov and Fabre}(2000)}]{kolobov_quantum_2000}
\bibinfo{author}{\bibfnamefont{M.~I.} \bibnamefont{Kolobov}} \bibnamefont{and}
  \bibinfo{author}{\bibfnamefont{C.}~\bibnamefont{Fabre}},
  \bibinfo{journal}{Phys. Rev. Lett.} \textbf{\bibinfo{volume}{85}},
  \bibinfo{pages}{3789} (\bibinfo{year}{2000}).

\bibitem[{\citenamefont{Sokolov et~al.}(2001)\citenamefont{Sokolov, Kolobov,
  Gatti, and Lugiato}}]{sokolov_quantum_2001}
\bibinfo{author}{\bibfnamefont{I.~V.} \bibnamefont{Sokolov}},
  \bibinfo{author}{\bibfnamefont{M.~I.} \bibnamefont{Kolobov}},
  \bibinfo{author}{\bibfnamefont{A.}~\bibnamefont{Gatti}}, \bibnamefont{and}
  \bibinfo{author}{\bibfnamefont{L.~A.} \bibnamefont{Lugiato}},
  \bibinfo{journal}{Opt. Comm.} \textbf{\bibinfo{volume}{193}},
  \bibinfo{pages}{175} (\bibinfo{year}{2001}).

\bibitem[{\citenamefont{Lassen et~al.}(2007)}]{lassen_tools_2007}
\bibinfo{author}{\bibfnamefont{M.}~\bibnamefont{Lassen}} \bibnamefont{et~al.},
  \bibinfo{journal}{Phys. Rev. Lett.} \textbf{\bibinfo{volume}{98}},
  \bibinfo{pages}{083602} (\bibinfo{year}{2007}).

\bibitem[{\citenamefont{Treps et~al.}(2003)}]{treps_quantum_2003}
\bibinfo{author}{\bibfnamefont{N.}~\bibnamefont{Treps}} \bibnamefont{et~al.},
  \bibinfo{journal}{Science} \textbf{\bibinfo{volume}{301}},
  \bibinfo{pages}{940} (\bibinfo{year}{2003}).

\bibitem[{\citenamefont{Mosset et~al.}(2005)\citenamefont{Mosset, Devaux, and
  Lantz}}]{mosset_spatially_2005}
\bibinfo{author}{\bibfnamefont{A.}~\bibnamefont{Mosset}},
  \bibinfo{author}{\bibfnamefont{F.}~\bibnamefont{Devaux}}, \bibnamefont{and}
  \bibinfo{author}{\bibfnamefont{E.}~\bibnamefont{Lantz}},
  \bibinfo{journal}{Phys. Rev. Lett.} \textbf{\bibinfo{volume}{94}},
  \bibinfo{pages}{223603} (\bibinfo{year}{2005}).

\bibitem[{\citenamefont{Choi et~al.}(1999)\citenamefont{Choi, Vasilyev, and
  Kumar}}]{choi_noiseless_1999}
\bibinfo{author}{\bibfnamefont{S.-K.} \bibnamefont{Choi}},
  \bibinfo{author}{\bibfnamefont{M.}~\bibnamefont{Vasilyev}}, \bibnamefont{and}
  \bibinfo{author}{\bibfnamefont{P.}~\bibnamefont{Kumar}},
  \bibinfo{journal}{Phys. Rev. Lett.} \textbf{\bibinfo{volume}{83}},
  \bibinfo{pages}{1938} (\bibinfo{year}{1999}).

\bibitem[{\citenamefont{Jedrkiewicz et~al.}(2004)}]{jedrkiewicz_detection_2004}
\bibinfo{author}{\bibfnamefont{O.}~\bibnamefont{Jedrkiewicz}}
  \bibnamefont{et~al.}, \bibinfo{journal}{Phys. Rev. Lett.}
  \textbf{\bibinfo{volume}{93}}, \bibinfo{pages}{243601}
  (\bibinfo{year}{2004}).

\bibitem[{\citenamefont{Martinelli
  et~al.}(2003)}]{martinelli_experimental_2003}
\bibinfo{author}{\bibfnamefont{M.}~\bibnamefont{Martinelli}}
  \bibnamefont{et~al.}, \bibinfo{journal}{Phys. Rev. A}
  \textbf{\bibinfo{volume}{67}}, \bibinfo{pages}{023808}
  (\bibinfo{year}{2003}).

\bibitem[{\citenamefont{McCormick
  et~al.}(2007{\natexlab{a}})\citenamefont{McCormick, Marino, Boyer, and
  Lett}}]{mccormick_strong_2007}
\bibinfo{author}{\bibfnamefont{C.~F.} \bibnamefont{McCormick}},
  \bibinfo{author}{\bibfnamefont{A.~M.} \bibnamefont{Marino}},
  \bibinfo{author}{\bibfnamefont{V.}~\bibnamefont{Boyer}}, \bibnamefont{and}
  \bibinfo{author}{\bibfnamefont{P.~D.} \bibnamefont{Lett}}
  (\bibinfo{year}{2007}{\natexlab{a}}), \urlprefix\url{quant-ph/0703111}.

\bibitem[{\citenamefont{McCormick
  et~al.}(2007{\natexlab{b}})\citenamefont{McCormick, Boyer, Arimondo, and
  Lett}}]{mccormick_strong_2007-1}
\bibinfo{author}{\bibfnamefont{C.~F.} \bibnamefont{McCormick}},
  \bibinfo{author}{\bibfnamefont{V.}~\bibnamefont{Boyer}},
  \bibinfo{author}{\bibfnamefont{E.}~\bibnamefont{Arimondo}}, \bibnamefont{and}
  \bibinfo{author}{\bibfnamefont{P.~D.} \bibnamefont{Lett}},
  \bibinfo{journal}{Opt. Lett.} \textbf{\bibinfo{volume}{32}},
  \bibinfo{pages}{178} (\bibinfo{year}{2007}{\natexlab{b}}).

\bibitem[{\citenamefont{Kumar and Kolobov}(1994)}]{kumar_degenerate_1994}
\bibinfo{author}{\bibfnamefont{P.}~\bibnamefont{Kumar}} \bibnamefont{and}
  \bibinfo{author}{\bibfnamefont{M.~I.} \bibnamefont{Kolobov}},
  \bibinfo{journal}{Opt. Comm.} \textbf{\bibinfo{volume}{104}},
  \bibinfo{pages}{374} (\bibinfo{year}{1994}).

\bibitem[{\citenamefont{Boyer et~al.}(2007)\citenamefont{Boyer, McCormick,
  Arimondo, and Lett}}]{boyer_ultraslow_2007}
\bibinfo{author}{\bibfnamefont{V.}~\bibnamefont{Boyer}},
  \bibinfo{author}{\bibfnamefont{C.~F.} \bibnamefont{McCormick}},
  \bibinfo{author}{\bibfnamefont{E.}~\bibnamefont{Arimondo}}, \bibnamefont{and}
  \bibinfo{author}{\bibfnamefont{P.~D.} \bibnamefont{Lett}},
  \bibinfo{journal}{to be published in Phys. Rev. Lett.}
  (\bibinfo{year}{2007}), \urlprefix\url{quant-ph/0703173}.

\bibitem[{\citenamefont{Fabre et~al.}(2000)\citenamefont{Fabre, Fouet, and
  Ma\^itre}}]{fabre_quantum_2000}
\bibinfo{author}{\bibfnamefont{C.}~\bibnamefont{Fabre}},
  \bibinfo{author}{\bibfnamefont{J.~B.} \bibnamefont{Fouet}}, \bibnamefont{and}
  \bibinfo{author}{\bibfnamefont{A.}~\bibnamefont{Ma\^itre}},
  \bibinfo{journal}{Opt. Lett.} \textbf{\bibinfo{volume}{25}},
  \bibinfo{pages}{76} (\bibinfo{year}{2000}).

\bibitem[{\citenamefont{Brambilla et~al.}(2004)\citenamefont{Brambilla, Gatti,
  Bache, and Lugiato}}]{brambilla_simultaneous_2004}
\bibinfo{author}{\bibfnamefont{E.}~\bibnamefont{Brambilla}},
  \bibinfo{author}{\bibfnamefont{A.}~\bibnamefont{Gatti}},
  \bibinfo{author}{\bibfnamefont{M.}~\bibnamefont{Bache}}, \bibnamefont{and}
  \bibinfo{author}{\bibfnamefont{L.~A.} \bibnamefont{Lugiato}},
  \bibinfo{journal}{Phys. Rev. A} \textbf{\bibinfo{volume}{69}},
  \bibinfo{pages}{023802} (\bibinfo{year}{2004}).

\bibitem[{\citenamefont{Harris et~al.}(1994)\citenamefont{Harris, Hill,
  Tapster, and Vaughan}}]{harris_laser_1994}
\bibinfo{author}{\bibfnamefont{M.}~\bibnamefont{Harris}},
  \bibinfo{author}{\bibfnamefont{C.~A.} \bibnamefont{Hill}},
  \bibinfo{author}{\bibfnamefont{P.~R.} \bibnamefont{Tapster}},
  \bibnamefont{and} \bibinfo{author}{\bibfnamefont{J.~M.}
  \bibnamefont{Vaughan}}, \bibinfo{journal}{Phys. Rev. A}
  \textbf{\bibinfo{volume}{49}}, \bibinfo{pages}{3119} (\bibinfo{year}{1994}).

\bibitem[{\citenamefont{Molina-Terriza
  et~al.}(2007)\citenamefont{Molina-Terriza, Torres, and
  Torner}}]{molina-terriza_twisted_2007}
\bibinfo{author}{\bibfnamefont{G.}~\bibnamefont{Molina-Terriza}},
  \bibinfo{author}{\bibfnamefont{J.~P.} \bibnamefont{Torres}},
  \bibnamefont{and} \bibinfo{author}{\bibfnamefont{L.}~\bibnamefont{Torner}},
  \bibinfo{journal}{Nat. Phys.} \textbf{\bibinfo{volume}{3}},
  \bibinfo{pages}{305} (\bibinfo{year}{2007}).

\bibitem[{\citenamefont{Villar et~al.}(2005)\citenamefont{Villar, Cruz,
  Cassemiro, Martinelli, and Nussenzveig}}]{villar_generation_2005}
\bibinfo{author}{\bibfnamefont{A.~S.} \bibnamefont{Villar}},
  \bibinfo{author}{\bibfnamefont{L.~S.} \bibnamefont{Cruz}},
  \bibinfo{author}{\bibfnamefont{K.~N.} \bibnamefont{Cassemiro}},
  \bibinfo{author}{\bibfnamefont{M.}~\bibnamefont{Martinelli}},
  \bibnamefont{and}
  \bibinfo{author}{\bibfnamefont{P.}~\bibnamefont{Nussenzveig}},
  \bibinfo{journal}{Phys. Rev. Lett.} \textbf{\bibinfo{volume}{95}},
  \bibinfo{pages}{243603} (\bibinfo{year}{2005}).

\bibitem[{\citenamefont{Delaubert et~al.}(2006)}]{delaubert_tem[sub_2006}
\bibinfo{author}{\bibfnamefont{V.}~\bibnamefont{Delaubert}}
  \bibnamefont{et~al.}, \bibinfo{journal}{Phys. Rev. A}
  \textbf{\bibinfo{volume}{74}}, \bibinfo{pages}{053823}
  (\bibinfo{year}{2006}).

\end{thebibliography}

\end{document}